\documentclass[twocolumn,pra,superscriptaddress,floatfix,a4paper,accepted=2026-01-19]{quantumarticle}

\pdfoutput=1

\usepackage[colorlinks=true,citecolor=blue,linkcolor=blue]{hyperref} 
\usepackage{graphicx}
\usepackage{bm}
\usepackage{amsmath}
\usepackage{amssymb}

\DeclareGraphicsExtensions{.png,.jpg,.eps}
\usepackage{color}
\usepackage{xcolor}

\newcommand{\beq}{\begin{equation}}
\newcommand{\eeq}{\end{equation}}

\usepackage{mleftright} 

\newcommand{\figref}[1]{\mbox{Fig.~\ref{#1}}}

\newcommand{\secref}[1]{\mbox{Section~\ref{#1}}}

\newcommand{\appref}[1]{\mbox{Appendix~\ref{#1}}}
\renewcommand{\eqref}[1]{\mbox{Eq.~(\ref{#1})}}

\newcommand{\figpanel}[2]{Fig.~\hyperref[#1]{\ref*{#1}(#2)}}
\newcommand{\figpanels}[3]{Fig.~\hyperref[#1]{\ref*{#1}(#2)-(#3)}}
\newcommand{\figpanelNoPrefix}[2]{\hyperref[#1]{\ref*{#1}(#2)}}

\newcommand{\ket}[1]{|#1 \rangle}
\newcommand{\bra}[1]{\langle #1|}

\newcommand{\ketbra}[2]{|#1\rangle \langle #2|}

\newcommand{\abs}[1]{\mleft|#1\mright|}

\usepackage{siunitx}

\begin{document}

\author{Guangze Chen}
\email{guangze@chalmers.se}
\affiliation{Department of Microtechnology and Nanoscience, Chalmers University of Technology, 41296 Gothenburg, Sweden}

\author{Anton Frisk Kockum}
\email{anton.frisk.kockum@chalmers.se}
\affiliation{Department of Microtechnology and Nanoscience, Chalmers University of Technology, 41296 Gothenburg, Sweden}


\title{Scalable quantum simulator with an extended gate set in giant atoms}

\begin{abstract}

Quantum computation and quantum simulation require a versatile gate set to optimize circuit compilation for practical applications. However, existing platforms are often limited to specific gate types or rely on parametric couplers to extend their gate set, which compromises scalability. Here, we propose a scalable quantum simulator with an extended gate set based on giant-atom three-level systems, which can be implemented with superconducting circuits. Unlike conventional small atoms, giant atoms couple to the environment at multiple points, introducing interference effects that allow exceptional tunability of their interactions. By leveraging this tunability, our setup supports both CZ and iSWAP gates through simple frequency adjustments, eliminating the need for parametric couplers. This dual-gate capability enhances circuit efficiency, reducing the overhead for quantum simulation. As a demonstration, we showcase the simulation of spin dynamics in dissipative Heisenberg XXZ spin chains, highlighting the setup's ability to tackle complex open quantum many-body dynamics. Finally, we discuss how a two-dimensional extension of our system could enable fault-tolerant quantum computation, paving the way for a universal quantum processor.

\end{abstract}


\maketitle


\section{Introduction}


Scalable universal quantum simulators are powerful tools for exploring complex quantum systems, including many-body physics in condensed matter physics and open quantum many-body dynamics~\cite{Georgescu2014, Altman2021, Fauseweh2024}. A universal gate set for a quantum simulator or quantum computer can be realized using a complete set of single-qubit gates combined with an entangling two-qubit gate, such as iSWAP or CZ~\cite{PhysRevA.52.3457}. Most existing quantum simulators are optimized for implementing only specific two-qubit gates~\cite{Wintersperger2023, Fauseweh2024, Strohm2024}, limiting their versatility. Expanding the available gate set beyond the bare minimum facilitates more efficient quantum circuit compilation~\cite{Leymann2020, kalloor2024quantumhardwarerooflineevaluating, Ge2024}, reducing circuit depth and improving performance. Notably, having access to both iSWAP and CZ gates enables any Clifford operation to be performed using single-qubit gates and no more than two two-qubit gates~\cite{Krian2025}. Moreover, introducing tunable qubit decay would open the door to simulating open quantum many-body dynamics. However, achieving a larger gate set typically requires additional resources, such as parametric couplers~\cite{Abrams2020,Ganzhorn2020, Sete2021, Rigetti, Krian2025,PhysRevLett.123.210501,PhysRevA.102.062408,PhysRevApplied.16.024050,PhysRevApplied.10.034050}, other coupling elements~\cite{Lacroix2020, Sung2021, Zhang2024,  ZChen2025,PhysRevApplied.23.024059,lvb9-pfr3}, or complicated drive schemes~\cite{Wei2022, Nguyen2024,PhysRevLett.129.060501,PRXQuantum.5.020326,Evered2023,PhysRevLett.125.120504,PRXQuantum.5.020338}, which limit the feasibility of building large-scale quantum simulators.


In this article, we show how giant artificial atoms~\cite{Kockum2021} can be used to build a scalable quantum simulator for open quantum systems with both iSWAP and controlled-phase (CZ$_\varphi$) two-qubit gates in the gate set. Unlike traditional small atoms, which couple to their environment at a single point, giant atoms couple at multiple discrete points, often separated by wavelengths. The consequences of having multiple coupling points have been explored in many articles in recent years, both theoretically in, e.g., Refs.~\cite{Kockum2021, Kockum2014, Guo2017, Kockum2018, Gonzalez-Tudela2019, Guo2020, Guimond2020, Ask2020, Cilluffo2020, Wang2021, Du2021, Soro2022, Wang2022, Du2022, Du2022a, Terradas-Brianso2022, Soro2023, Du2023, PhysRevResearch.6.043222, Leonforte2024, Wang2024, Roccati2024, Gong2024, Du2025} and in experiments mostly using superconducting circuits, e.g., in Refs.~\cite{Gustafsson2014, Manenti2017, Satzinger2018, Bienfait2019, Andersson2019, Kannan2020, Bienfait2020, Andersson2020, Vadiraj2021, Wang2022a, Joshi2023, Hu2024}. The main point is that the multiple coupling points produce interference effects, enabling frequency-dependent control of relaxation rates~\cite{Kockum2014, Vadiraj2021} and qubit-qubit interaction strengths~\cite{Kockum2018}. This tunability makes giant atoms well-suited for implementing diverse gate operations without overhead like parametric couplers. For instance, iSWAP gates and tunable decays can be achieved in giant-atom systems by tuning qubit frequencies~\cite{Kannan2020}, enabling applications such as the Trotterized simulation of quantum Zeno dynamics in open quantum systems~\cite{Chen2025}. However, an extended gate set including CZ operations has not been studied previously in giant-atom setups.


Here, to enable the execution of CZ$_\varphi$ gates in addition to iSWAP gates, we introduce three-level giant atoms with additional coupling points to eliminate unwanted interactions and reduce errors. We demonstrate the ability of our simulator built from such systems to simulate the dynamics of the dissipative XXZ model, illustrating the advantages of an extended gate set in reducing simulation errors. Additionally, we propose a two-dimensional extension of our quantum simulator, which enables the execution of long-range two-qubit operations. This extension further allows the implementation of surface codes for quantum error correction~\cite{Fowler2012}, thereby supporting fault-tolerant quantum computation and positioning our setup as a potential universal quantum processor. 

The rest of this article is organized as follows. In \secref{sec2}, we put forward a setup with two three-level giant atoms as the building block of our scalable quantum simulator. We demonstrate how this building block can realize both iSWAP and CZ gates, as well as the more general $R_\text{XY}$ and controlled-phase (CZ$_\varphi$) gates, through simple frequency tuning of the giant atoms. We then analyze the average gate fidelity of these gates under realistic noise conditions, showing that state-of-the-art techniques yield fidelities of $\geq 98.8\%$ for both iSWAP and CZ operations; these fidelities can quite easily be increased by coupling the giant atoms more strongly to the waveguide. Then, in \secref{sec3}, we combine such building blocks into a scalable simulator architecture where the two-giant-atom structure is repeated to form a one-dimensional chain. We provide a protocol for tuning the atomic frequencies to implement nearest-neighbor $R_\text{XY}$ and CZ$_\varphi$ gates across this simulator, demonstrating the scalability of the extended gate set. 

To highlight the simulator's potential, we showcase in \secref{sec4} its simulation of spin dynamics in a dissipative XXZ spin chain~\cite{Mi2024, Rosenberg2024, PhysRevLett.106.217206, PhysRevLett.106.220601}, illustrating its ability to handle complex open quantum many-body system. To further enhance the capability of our simulator to execute long-range two-qubit gates, we propose in \secref{sec5} a two-dimensional extension of it. We discuss how long-range two-qubit gates can be operated in such a setup by tuning the frequencies of the giant atoms. In particular, this extension allows to perform surface codes, enabling fault-tolerant quantum computation, and offering our setup as a universal quantum processor. We provide a discussion on the physical realization of our setup and the influence of non-Markovian effects in \secref{sec6}, followed by concluding remarks in \secref{sec7}. Some additional details on the error analysis for the quantum simulation in \secref{sec4} are provided in \appref{sec_err}.


\section{Two-qubit gates with giant atoms} \label{sec2}

\begin{figure}
\center
\includegraphics[width=\linewidth]{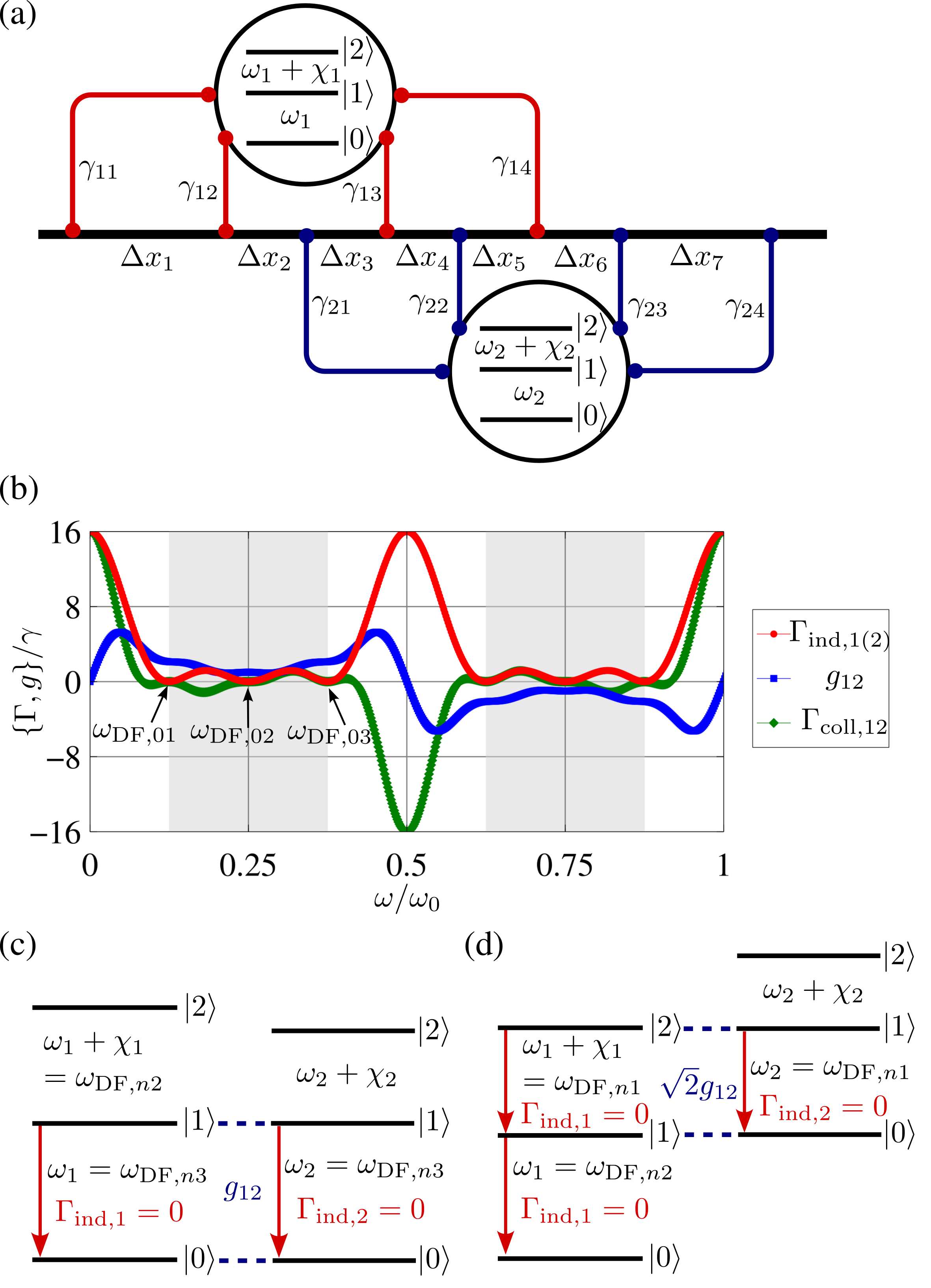}
\caption{A two-giant-atom setup for performing both iSWAP and CZ gates. 
(a) Sketch of the setup. The two giant atoms, with frequencies $\omega_{1,2}$ and detunings $\chi_{1,2}$, are coupled to the waveguide (black line) at multiple points with different coupling strengths $\gamma_{kn}$ and spacings $\Delta x_n$. The coupling points are organized in a braided fashion.
(b) Frequency dependence of the individual decay rates $\Gamma_{1,2}$, inter-atomic coupling strength $g_{12}$, and collective decay rates $\Gamma_{\rm coll,12}$ of the giant atoms. 
(c,d) The protocol to perform (c) an iSWAP gate and (d) a CZ gate in this setup.}
\label{fig1}
\end{figure}

We begin by studying a two-giant-atom setup that allows us to perform both iSWAP and CZ gates by simple frequency tuning. This setup serves as the building block for our scalable giant-atom-based simulator. The giant atoms we consider are $\Xi$-type three-level systems with an anharmonicity $\chi_{k} = \omega_{12,k} - \omega_{01,k} <0$, where $\omega_{ij,k}$ is the transition frequency between states $\ket{i}$ and $\ket{j}$ in atom $k$; for simplicity, we use in the following the notation $\omega_{01,k} \equiv \omega_k$. Such negative anharmonicity and ladder-type level structure are typical for superconducting transmon qubits~\cite{Koch2007}, which is the most common platform for experiments on giant atoms so far. The giant atoms are coupled to a waveguide at multiple spatially separated points each (with coordinate $x_{kn}$ for the $n$th coupling point of atom $k$), and these coupling points are organized in a braided fashion, as shown in \figpanel{fig1}{a}.  

Due to the coupling to the waveguide, the two atoms acquire individual decay rates $\Gamma_{\text{ind},k}(\omega)$, a coherent interaction with strength $g_{jk}(\omega)$ between atoms $j$ and $k$, and a collective decay rate $\Gamma_{\text{coll},jk}(\omega)$ for atoms $j$ and $k$. The dependence on the frequency $\omega$ of the transition that these rates hold for is a consequence of the interference effects arising from the multiple coupling points of giant atoms. Assuming Markovianity, i.e., that the travel time between coupling points is negligible, the decay rates and interaction strengths are given by~\cite{Kockum2018}
\beq \label{eq1}
\begin{aligned}
\Gamma_{\text{ind},k}(\omega)=\sum_{n=1}^{N_k}\sum_{m=1}^{N_k}\sqrt{\gamma_{kn}\gamma_{km}}\cos\phi_{kn,km}(\omega),\\
g_{jk}(\omega)=\sum_{n=1}^{N_j}\sum_{m=1}^{N_k}\frac{\sqrt{\gamma_{jn}\gamma_{km}}}{2}\sin\phi_{jn,km}(\omega),\\
\Gamma_{\text{coll},jk}(\omega)=\sum_{n=1}^{N_j}\sum_{m=1}^{N_k}\sqrt{\gamma_{jn}\gamma_{km}}\cos\phi_{jn,km}(\omega),
\end{aligned}
\eeq
where $N_k$ is the number of coupling points of atom $k$, $\gamma_{kn}$ is the coupling strength at the $n$th coupling point of atom $k$, $\phi_{jn,km}(\omega) = \omega\Delta x_{jn,km}/v$ is the accumulated phase between the coupling points, $\Delta x_{jn,km} = \abs{x_{jn} - x_{km}}$ the distance between the coupling points, and $v$ the speed of light in the waveguide, which we assume to have linear dispersion. We note that using nonlinear structured waveguides~\cite{Soro2023} would not qualitatively change our results.

The interference effects in the giant atoms result in decoherence-free frequencies where the atomic decay into the waveguide vanishes.
To illustrate this, we consider that all the coupling strengths are identical ($\gamma$), and $\Delta x_1/2=\Delta x_7/2=\Delta x_2=\Delta x_3=\Delta x_4=\Delta x_5=\Delta x_6=\Delta x$ in \figpanel{fig1}{a}. This leads to a periodic dependence of $\Gamma_{\text{ind},k}(\omega)$, $g_{jk}(\omega)$, and $\Gamma_{\text{coll},jk}(\omega)$ on $\omega$, with a periodicity of $\omega_0=2\pi v/\Delta x$. Notable are the decoherence-free frequencies $\omega_{\text{DF},nm}=(n+m/8)\omega_0$ $(n\in\mathcal{N}, m=1,2,3,5,6,7)$, at which the interaction $g_{12}$ is non-zero, as shown in \figpanel{fig1}{b}. These frequencies are essential for performing high-fidelity two-qubit gates. 

Moreover, the individual decay rates are minimal within the frequency ranges $[\omega_{\text{DF},n1},\omega_{\text{DF},n3}]$ and $[\omega_{\text{DF},n5},\omega_{\text{DF},n7}]$, as highlighted in the grey regions in \figpanel{fig1}{b}. This behavior results from having four coupling points for each giant atom; the regions can be extended by having more coupling points. As we will see in the following, this property reduces qubit decay during gate operations and provides a broad operational frequency range. To enable both iSWAP and CZ gates with high fidelity, we set $\chi_1=-\chi=-\omega_0/8$.

We now show how different gates can be performed with this setup. The first thing to notice is that when the two atoms are at different decoherence-free frequencies, they will have negligible coupling due to the detuning between them, and thus the system remains in a steady state. On top of this, single-qubit decay can be implemented by tuning the qubit's frequency to a non-decoherence-free value, allowing controlled dissipation. This feature is particularly beneficial for simulating open quantum systems, as explored in \secref{sec4}.


\subsection{iSWAP gate with giant atoms}

Decoherence-free interactions between braided giant atoms facilitate the realization of the iSWAP gate~\cite{Kannan2020}. The action of an iSWAP gate leaves the states $\ket{00}$ and $\ket{11}$ unchanged, while the states $\ket{01}$ and $\ket{10}$ are swapped and acquire a phase factor $i$. This outcome is achieved in our setup by setting $\omega_1=\omega_2=\omega_{\text{DF},nm}$ such that the $\ket{0} \leftrightarrow \ket{1}$ transitions of the two atoms are resonant [\figpanel{fig1}{c}].

\begin{figure*}
\center
\includegraphics[width=\linewidth]{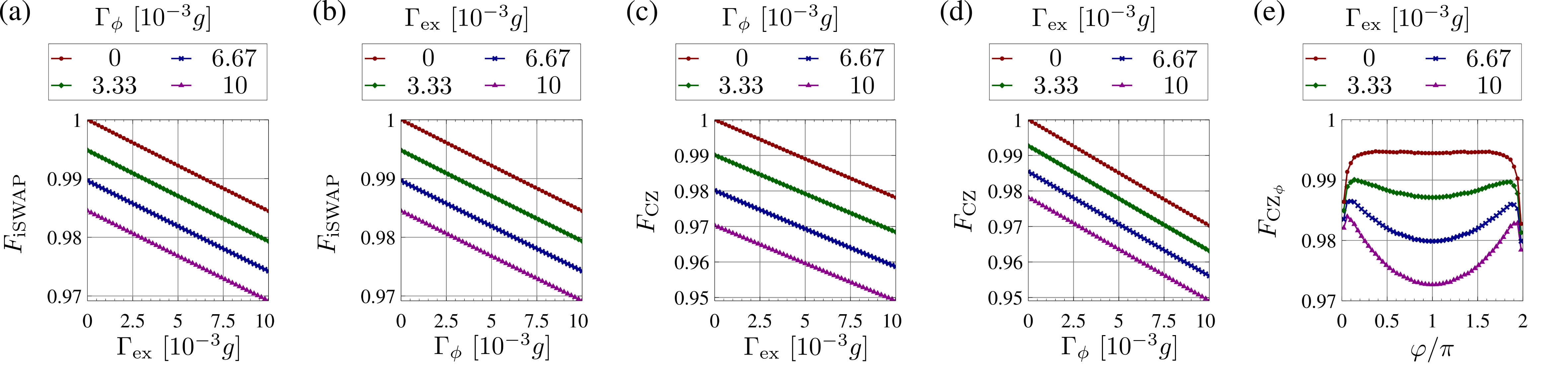}
\caption{Average process fidelity of two-qubit (a,b) iSWAP (c,d) CZ and (e) CZ$_\varphi$ gates performed with the setup in \figpanel{fig1}{a}, as a function of qubit decay rate $\Gamma_\text{ex}$ and dephasing rate $\Gamma_\phi$. Here, $g\approx2.1\gamma$ is the qubit-qubit coupling strength used in the gates.}
\label{fig2}
\end{figure*}

We now analyze the fidelity of the giant-atom iSWAP gate in realistic cases, when the qubits are subject to decay and dephasing. To be concrete, we consider the case of transmon qubits, where the decay and dephasing of the $\ket{2}$ state are twice those for the $\ket{1}$ state~\cite{Koch2007}. Additionally, for transmon qubits, a direct decay from $|2\rangle$ to $|0\rangle$ is prevented due to a vanishingly small matrix element connecting the two states. Thus, the Lindblad dissipators for qubit decay and dephasing are
\begin{align}
L_- &=\sqrt{\Gamma_{\rm ind} + \Gamma_{\rm ex}} \mleft( \ketbra{1}{0} + \sqrt{2}\ketbra{2}{1} \mright), \\
L_\phi &=\sqrt{2\Gamma_\phi} \mleft( \ketbra{1}{1} + 2 \ketbra{2}{2} \mright) ,
\end{align}
respectively, where $\Gamma_\text{ex}$ is the extra decay rate of the qubit to environments other than the waveguide, and $\Gamma_\phi$ is the qubit's dephasing rate. For simplicity, we further assume that the two qubits have the same decay and dephasing rates. We also neglect the effect of tuning the atoms in and out of the conditions enabling the gate; we return to that topic in \secref{sec4}.

Taking into account these decoherence processes, we show in \figpanels{fig2}{a}{b} the effect of decay and dephasing on the average process fidelity~\cite{Gilchrist2005} of the iSWAP gate,
\begin{equation}
F_\text{iSWAP}=\mleft[\text{tr}\mleft(\sqrt{\sqrt{\Phi}\Phi_0\sqrt{\Phi}}\mright)\mright]^2 ,
\end{equation}
where $\Phi$ and $\Phi_0$ are the Choi matrices~\cite{Choi1975} of our gate and the perfect iSWAP gate, respectively. The Choi matrix $\Phi$ of a process $\mathcal{E}$ is defined as
\beq
\Phi = \sum_n \ket{n}\bra{n} \otimes \mathcal{E}(\ket{n}\bra{n}).
\eeq
where $n$ runs over the basis states of the computational subspace. We observe an expected linear dependence~\cite{PhysRevLett.129.150504} of the process fidelity on both the decay and dephasing: $F_\text{iSWAP}\approx1-1.57\Gamma_\text{ex}/g-1.57\Gamma_\phi/g$, where $g$ is the qubit-qubit coupling. This means the average gate fidelity of the iSWAP gate is given by $F_{\text{ave, iSWAP}}\approx1-1.26\Gamma_\text{ex}/g-1.26\Gamma_\phi/g$, in agreement with previous results~\footnote{The average gate fidelity and the process fidelity are related as~\cite{Gilchrist2005}: $F_\text{ave}=1-dF/(d+1)$ where $F_\text{ave}$ and $F$ are the average gate and process fidelities, and $d$ is the dimension of the computational space. For two-qubit iSWAP and CZ gates, we take $d=4$. The average gate fidelity of a two-qubit iSWAP gate is $F_{\text{ave, iSWAP}}\approx1-0.8\Gamma_\text{ex}\tau-0.8\Gamma_\phi\tau$, where $\tau$ is the gate time~\cite{PhysRevLett.129.150504}. For two qubits coupled with interaction strength $g$, the gate time is $\tau=\pi/(2g)$. Combining these equations, we find that it agrees with the equation in the main text.}. Assuming that the iSWAP gate is performed with $\omega_1=\omega_2=\omega_{\text{DF},n3}$, which yields $g\approx2.1\gamma$, and taking an experimentally accessable value of $\gamma/ (2\pi) = \qty{2}{\mega\hertz}$, $\Gamma_{\rm ex} = \qty{0.02}{\mega\hertz}\approx 0.76 \cdot 10^{-3}g$ and $\Gamma_\phi = \qty{0.05}{\mega\hertz} \approx 1.89 \cdot 10^{-3}g$~\cite{annurev_qubits, Place2021, Somoroff2023, Kim2023, Biznarova2023, Kono2024, Bal2024}, the average gate fidelity is $\qty{99.67}{\percent}$. To achieve even higher fidelity, we can increase the qubit-waveguide coupling $\gamma$ and consequently the qubit-qubit coupling $g$, such that the gate time becomes smaller; for example, with $\gamma/ (2\pi) = \qty{4}{\mega\hertz}$, an average gate fidelity of $\qty{99.83}{\percent}$ can be achieved.

As the iSWAP gate in our setup is performed via the XY interactions between the giant atoms, an $R_\text{XY}(\theta)$ gate can in general be performed with our setup. This can be achieved by setting the interacting time $\tau=\theta/g$ for $g>0$ and $\tau=(2\pi-\theta)/|g|$ for $g<0$. In particular, the iSWAP gate is $R_\text{XY}(\pi/2)$.


\subsection{CZ and CZ$_\varphi$ gates with giant atoms}

Our setup also allows the implementation of CZ and CZ$_\varphi$ gates between giant atoms. The CZ gate, which adds a phase of $\pi$ to the $\ket{11}$ state of the two-qubit system and leaves all other states unchanged, can be achieved by bringing the population of the $\ket{11}$ state to $\ket{02}$ or $\ket{20}$ and back~\cite{Strauch2003}. This process requires a resonant transition between these two states. In our setup, this resonance is achieved by having $\omega_2=\omega_1+\chi_1$ (or $\omega_1=\omega_2+\chi_2$) such that there is an interaction between the $\ket{11}$ and $\ket{20}$ (or $\ket{02}$) states. To ensure that at the same time no extra decay happens for the involved levels, we can take $\omega_1=\omega_{\text{DF},n2}$ and $\omega_2=\omega_{\text{DF},n1}$ [\figpanel{fig1}{d}]; this choice yields a qubit-qubit coupling $g\approx2.1\gamma$. The coupling between the corresponding levels is then $\sqrt{2}g$, where the factor of $\sqrt{2}$ originates from the fact that transmons are close to harmonic oscillators~\cite{Kockum2014, Koch2007}.

The average process fidelity of the CZ gate implemented in this way is shown in \figpanels{fig2}{c}{d}. Due to the fact that CZ involves higher levels with higher decay and dephasing rates, and that it takes $\sqrt{2}$ longer time than iSWAP for the same coupling strength $g$, the process fidelity is lower than for iSWAP: $F_\text{CZ}\approx1-2.19\Gamma_\text{ex}/g-2.97\Gamma_\phi/g$. Correspondingly, the average gate fidelity is $F_{\text{ave, CZ}}\approx1-1.75\Gamma_\text{ex}/g-2.34\Gamma_\phi/g$. The average gate fidelity shows a stronger dependence on dephasing, in agreement with previous results~\cite{Abad2025, footnote_CZ}. Taking a typical value of $\gamma/ (2\pi) = \qty{2}{\mega\hertz}$, $\Gamma_\text{ex} = \qty{0.02}{\mega\hertz}$ and $\Gamma_\phi = \qty{0.05}{\mega\hertz}$, the average gate fidelity is $\qty{99.42}{\percent}$ for our setup. The gate fidelity can be improved by increasing the qubit-waveguide coupling $\gamma$; for example, with $\gamma/ (2\pi) = \qty{4}{\mega\hertz}$, the average gate fidelity becomes $\qty{99.71}{\percent}$.

We can also perform a generalized controlled-phase gate CZ$_\varphi$ by introducing a detuning from the resonance condition used in the CZ gate. Specifically, setting $\omega_2=\omega_1+\chi_1+\Delta$ produces the following Hamiltonian in the subspace spanned by $\ket{11}$ and $\ket{20}$:
\beq
H_{\rm CZ\varphi}=
\begin{pmatrix}
   \Delta/2  & \sqrt{2}g \\
   \sqrt{2}g  & -\Delta/2
\end{pmatrix} ,
\eeq
where $\sqrt{2}g$ is the coupling between $\ket{11}$ and $\ket{20}$. Starting in $\ket{11}$, the shortest evolution time that returns the state to $\ket{11}$ (up to a phase) is $\tau=\pi/g'$, where $g'=\sqrt{2g^2+\Delta^2/4}$. This yields
\beq
\exp(iH_{\rm CZ\varphi}\pi/g')\ket{11}=\exp(i\varphi)\ket{11}
\eeq
with a $\Delta$-dependent phase $\varphi$~\cite{Lacroix2020}:
\beq \label{eq2}
\varphi=\pi\left(1+\frac{\Delta}{\sqrt{8g^2+\Delta^2}}\right).
\eeq
The detuning $\Delta$ in qubit 2 with respect to the decoherence-free frequency results in the decay of its $|1\rangle$ level to the waveguide, which depends on $\Delta/\omega_0$. The process fidelity of the CZ$_\varphi$ gate is thus influenced by both the shorter gate time and the decay into the waveguide induced by the detuning. 

To analyze the process fidelity for the CZ$_\varphi$ gate, we consider a realistic situation of $\chi/(2\pi)=\qty{200}{\mega\hertz}$, which yields $\omega_0/(2\pi)=\qty{1.6}{\giga\hertz}=800\gamma$; we also fix the dephasing rate to a value of $\Gamma_\phi= \qty{0.05}{\mega\hertz}$ without loss of generality. The process fidelity for different $\varphi$ and $\Gamma_\text{ex}$ is shown in \figpanel{fig2}{e}. We observe that, for $|\varphi-\pi|<\varphi_c$, where $\varphi_c\approx0.9\pi$, the process fidelity is sensitive to $\Gamma_\text{ex}$, while for $\varphi$ close to 0 and $2\pi$ it is not. This is because to have $\varphi$ close to 0 and $2\pi$, the detuning $\Delta$ in \eqref{eq2} will be large, resulting in a large decay into the waveguide that dominates over the intrinsic qubit decay $\Gamma_\text{ex}$. For smaller $|\varphi-\pi|$, the intrinsic qubit decay $\Gamma_\text{ex}$ dominates over decay into the waveguide, and thus the fidelity is mostly influenced by $\Gamma_\text{ex}$ and the gate time $\tau$. In particular, when $|\varphi-\pi|$ increases, the gate time decreases, and a higher fidelity is obtained. Unlike the CZ gate, whose fidelity increases with $\gamma$ due to the shorter gate time, the fidelity of the CZ$_\varphi$ gate does not increase monotonically with $\gamma$. While a larger $\gamma$ speeds up the gate, it also enhances decay of the detuned qubit into the waveguide, which lowers the fidelity. As this decay increases with $\Delta/\omega_0$, a larger $\omega_0$ reduces it and lowers $\gamma_c$, allowing high-fidelity CZ$_\varphi$ gates over a wider range of $\varphi$.

We finally note that, taking $\omega_1=\omega_{\text{DF},n3}$, different two-qubit operations can be achieved by solely tuning $\omega_2$ within $[\omega_{\text{DF},n1},\omega_{\text{DF},n3}]$: to have no evolution, $\omega_2=\omega_{\text{DF},n1}$; to have CZ, $\omega_2=\omega_{\text{DF},n2}$, and to have iSWAP, $\omega_2=\omega_{\text{DF},n3}$. As the individual decay rates within this frequency range are small, highly tunable two-qubit gates with high fidelity can be realized here. We also note that negative couplings $g<0$ can be achieved in our setup by switching to the frequency regime of $[\omega_{\text{DF},n5},\omega_{\text{DF},n7}]$ --- this allows to perform inverse operations of $R_\text{XY}(\theta)$. The setup's versatility and tunability establish it as a robust building block for scalable quantum simulators.


\section{Scalable giant-atom-based quantum simulator} \label{sec3}

The tunability of giant atoms enables controlled single-qubit decay, iSWAP gates, and CZ gates via coupling to a waveguide, forming an extended gate set. All these operations can be realized using the simple structure depicted in \figpanel{fig1}{a}, making the setup inherently scalable toward a many-body quantum simulator for open quantum systems. In this section, we demonstrate how such a scalable simulator can be constructed, in a configuration optimized for nearest-neighbor two-qubit iSWAP and CZ gates. Additionally, we note that alternative architectures, such as one supporting all-to-all tunable couplings, are feasible using a similar approach as in Ref.~\cite{Chen2025}.


\subsection{Setup}

\begin{figure}
\center
\includegraphics[width=\linewidth]{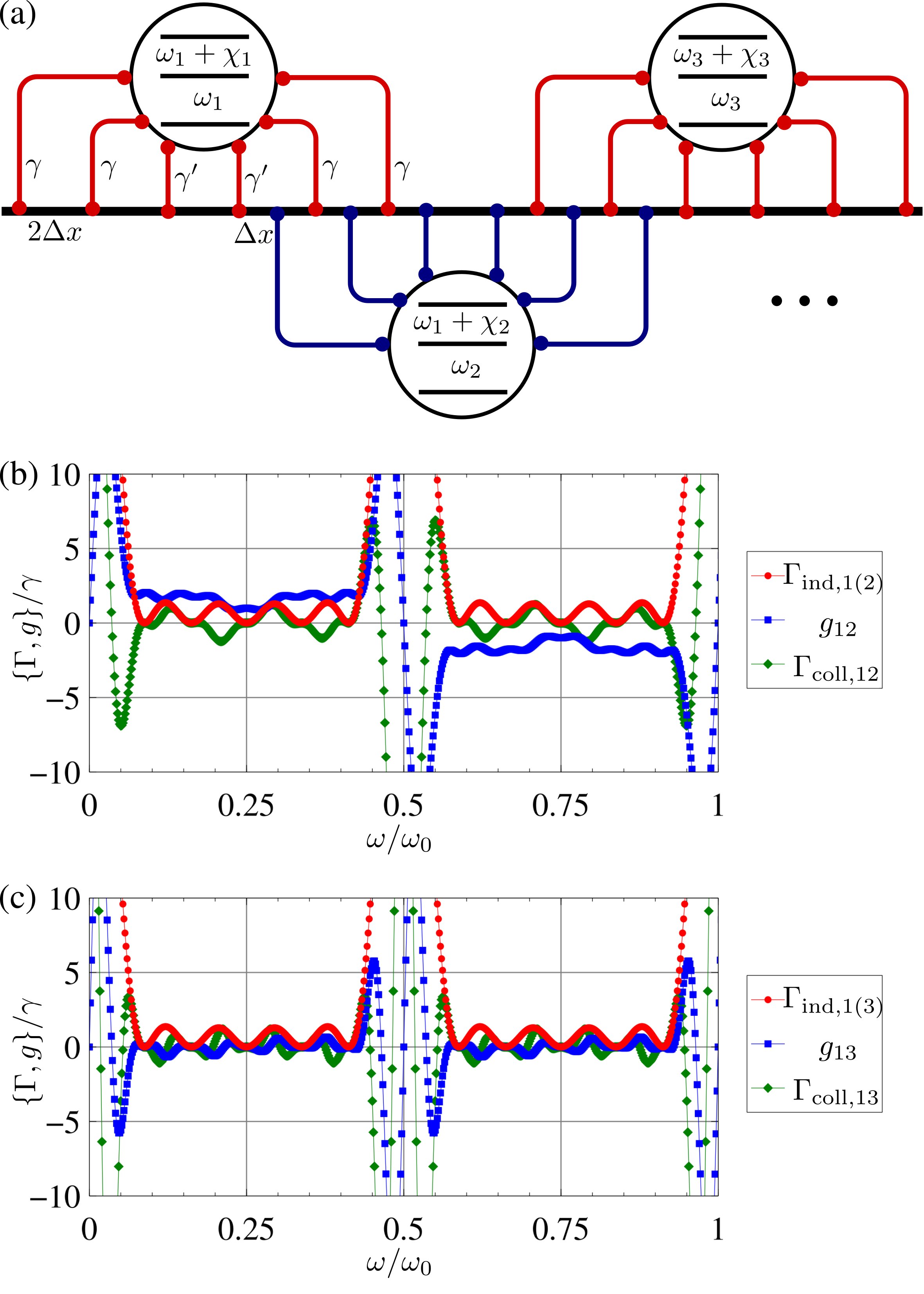}
\caption{Scalable giant-atom-based quantum simulator. 
(a) The architecture of the quantum simulator, where neighboring giant atoms are coupled to the waveguide in a braided configuration. 
(b,c) The frequency dependence of individual decay rates $\Gamma_\text{ind}$, coupling strength $g$, and collective decay rates $\Gamma_\text{coll}$ for (b) neighboring and (c) non-neighboring giant atoms.}
\label{fig3}
\end{figure}
 
The architecture of the scalable simulator is shown in \figpanel{fig3}{a}, where neighboring giant atoms are coupled to the waveguide in a braided fashion. Compared to \figpanel{fig1}{a}, two more coupling points per atom have been added to create more decoherence-free frequencies for the performance of two-qubit operations between different neighbors. To minimize individual atom decay within the operational frequency range, the coupling strength at the middle connection points of each atom is set slightly higher than that of the outer ones: $\gamma' = 1.4\gamma$. 

The frequency dependence of the coupling strength $g$, individual decay rates $\Gamma_\text{ind}$, and collective decay rates $\Gamma_\text{coll}$ between neighboring and non-neighboring giant atoms, is depicted in \figpanels{fig3}{b}{c}. A set of decoherence-free frequencies $\omega_{\text{DF},nm}\quad (n\in\mathcal{N}, m=1,\dots,10)$ allows coupling between neighboring atoms ($g_{12}\neq0$), while suppressing unwanted coupling between non-neighboring atoms ($g_{13}=0$). This feature is crucial for ensuring that only intended qubits interact during gate operations. The frequency regime $[\omega_{\text{DF},n1},\omega_{\text{DF},n5}]$, which includes five decoherence-free frequencies, is ideal for operating the two-qubit gates due to minimal individual decay in this range.

The qubits in the simulator are arranged with odd-site qubits placed at fixed frequencies $\omega_{4k-3}=\omega_{\text{DF},n2}$, $\omega_{4k-1}=\omega_{\text{DF},n5}$ ($k\in\mathcal{N}$), where $n$ is chosen such that $\omega_{\text{DF},n2}$ and $\omega_{\text{DF},n5}$ lie in the optimal frequency regime for the qubits. The even-site qubits are tunable, enabling diverse gate operations with their neighbors by adjusting their frequencies. This design minimizes errors from decoherence during frequency tuning, since only half of the qubits require tunability, and then only in a limited range.


\subsection{Performing gates}

We now discuss how different qubit operations can be executed in the scalable simulator. To maintain a steady state of the simulator, we can set $\omega_{2k}=\omega_{\text{DF},n3}$
such that all atoms are decoherence-free, while neighboring atoms are detuned and will not couple to each other. On top of this, single-qubit decay can be achieved by tuning the targeted qubit to a non-decoherence-free frequency.

\begin{figure}[t!]
\center
\includegraphics[width=\linewidth]{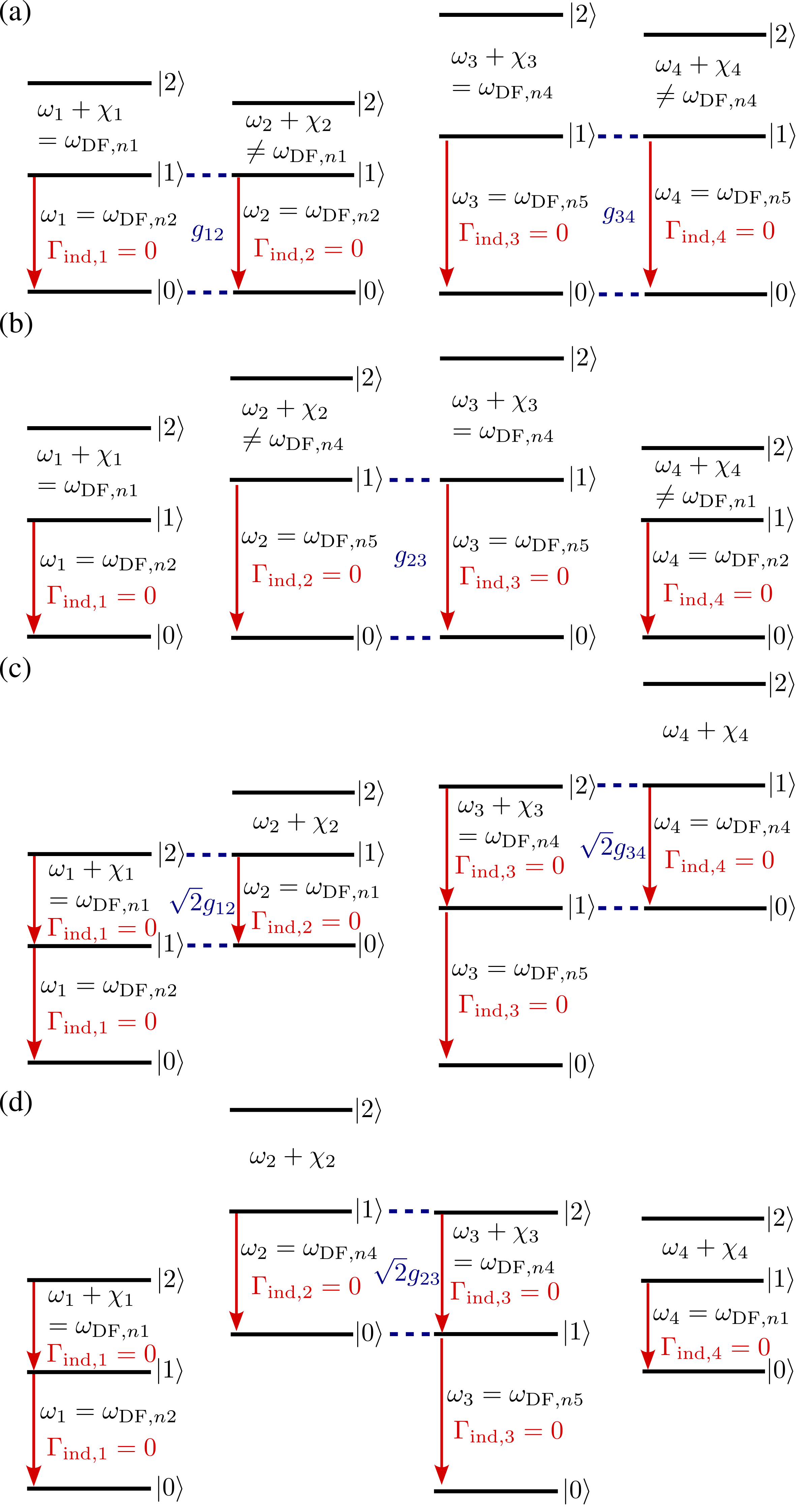}
\caption{Protocol to perform different two-qubit operations on the giant-atom-based simulator. The qubits' frequencies are tuned to achieve: (a) $R_\text{XY}(\theta)$ between qubits $2k-1$ and $2k$, (b) $R_\text{XY}(\theta)$ between qubits $2k$ and $2k+1$, (c) CZ between qubits $2k-1$ and $2k$ and (d) CZ between qubits $2k$ and $2k+1$.}
\label{fig4}
\end{figure}

To perform iSWAP gates, and in general $R_\text{XY}(\theta)$ gates, between neighboring atoms, there are two possibilities: (i) $R_\text{XY}(\theta)$ between qubits $2k-1$ and $2k$ and (ii) $R_\text{XY}(\theta)$ between qubits $2k$ and $2k+1$. To achieve (i), we set ($k\in\mathcal{N}$) [\figpanel{fig4}{a}]
\beq
\omega_{4k-2}=\omega_{\text{DF},n2},\quad
\omega_{4k}=\omega_{\text{DF},n5},
\eeq
such that qubits $4k-3$ and $4k-2$ are coupled with a strength of $g'_1\approx1.79\gamma$, qubits $4k-1$ and $4k$ are coupled with a strength of $g'_2\approx2.05\gamma$, and qubits $2k$ and $2k+1$ are detuned and thus are decoupled. Similarly, (ii) can be achieved with ($k\in\mathcal{N}$) [\figpanel{fig4}{b}]
\beq
\omega_{4k-2}=\omega_{\text{DF},n5},\quad
\omega_{4k}=\omega_{\text{DF},n2}.
\eeq

Similar to iSWAP gates, there are also two possibilities to perform CZ gates: (i) CZ between qubits $2k-1$ and $2k$ and (ii) CZ between qubits $2k$ and $2k+1$. To achieve (i), we set [\figpanel{fig4}{c}]
\beq
\begin{aligned}
\omega_{4k-2}=\omega_{\text{DF},n1},\quad\omega_{4k}=\omega_{\text{DF},n4},
\end{aligned}
\eeq
and similarly for (ii) we set [\figpanel{fig4}{d}]
\beq
\begin{aligned}
\omega_{4k-2}=\omega_{\text{DF},n4},\quad\omega_{4k}=\omega_{\text{DF},n1}.
\end{aligned}
\eeq
On top of this, to perform CZ$_\varphi$ gates, a detuning $\Delta$ given by \eqref{eq2} can be added to the qubits on even sites.

We have thus demonstrated an architecture for a scalable quantum simulator with an extended gate set of $R_\text{XY}(\theta)$ and CZ$_\varphi$ gates with giant atoms. Here, the interference effects mediated by the waveguide not only enable different qubit operations with simple frequency tuning, but also eliminate unwanted couplings between non-neighboring qubits. The extended gate set reduces circuit depth for simulation tasks requiring a combination of these operations, and thus reduces simulation errors. Furthermore, the simulator’s controllable qubit decay is uniquely suited for simulating open quantum dynamics. We illustrate these benefits in the next section with a concrete example of quantum simulation of the dynamics of an open quantum system.


\section{Application in quantum simulation} \label{sec4}

The extended gate set of our giant-atom-based simulator makes it versatile for simulating a broad range of open quantum dynamics, such as the dynamics of dissipative XXZ chains~\cite{Mi2024, Rosenberg2024,PhysRevLett.106.217206,PhysRevLett.106.220601} and the quantum contact process~\cite{Chertkov2023,PhysRevLett.123.100604,PhysRevLett.116.245701}. In this section, we illustrate the versatility of our simulator by showcasing the simulation of a dissipative XXZ model for $N$ spins,
\beq \label{eq_model_sim}
H=\sum_{k=1}^{N-1}\mleft[ J \mleft( \sigma_k^x\sigma_{k+1}^x+\sigma_k^y\sigma_{k+1}^y \mright)+J_z\sigma_k^z\sigma_{k+1}^z \mright],
\eeq
where $J$ and $J_z$ are coupling strengths and we add dissipation on the last site: $L=\sqrt{\Gamma}\sigma_N^-$. Here $\sigma^{x,y,z}$ are Pauli matrices and $\sigma^-=\sigma^x-i\sigma^y$.

To simulate the dynamics given by the above model,  we employ the Trotter-Suzuki decomposition~\cite{SUZUKI1990319, Kliesch2011}
\beq
\exp \mleft( \mathcal{L} t \mright) = \mleft[ \prod_{j=1}^n \exp\mleft( \mathcal{L}_j t/l \mright) \mright]^l + O \mleft( \frac{t^2}{l} \mright) ,
\label{eq:TrotterSuzuki}
\eeq
where $\mathcal{L}[\rho]=-i[H,\rho]+L\rho L^\dag-\frac{1}{2}(L^\dag L\rho+\rho L^\dag L)$ is the Liouvillian superoperator governing the dynamics of this model, $l$ is the number of Trotter steps, and $\mathcal{L}=\sum_j \mathcal{L}_j$. We divide $\mathcal{L}$ into components $\mathcal{L}_{1,2,3,4}[\rho]=-i[H_{1,2,3,4},\rho]$ and $\mathcal{L}_5[\rho]=L\rho L^\dag-\frac{1}{2}(L^\dag L\rho+\rho L^\dag L)$, where
\beq
\begin{aligned}
&H_1=J\sum_{k=1}^{N/2}\mleft(\sigma_{2k-1}^x\sigma_{2k}^x+\sigma_{2k-1}^y\sigma_{2k}^y\mright) ,\\
&H_2=J_z\sum_{k=1}^{N/2}\sigma_{2k-1}^z\sigma_{2k}^z ,\\
&H_3=J\sum_{k=1}^{N/2-1}\mleft(\sigma_{2k}^x\sigma_{2k+1}^x+\sigma_{2k}^y\sigma_{2k+1}^y\mright) ,\\
&H_4=J_z\sum_{k=1}^{N/2-1}\sigma_{2k}^z\sigma_{2k+1}^z.
\end{aligned}
\eeq

\begin{figure}
\center
\includegraphics[width=\linewidth]{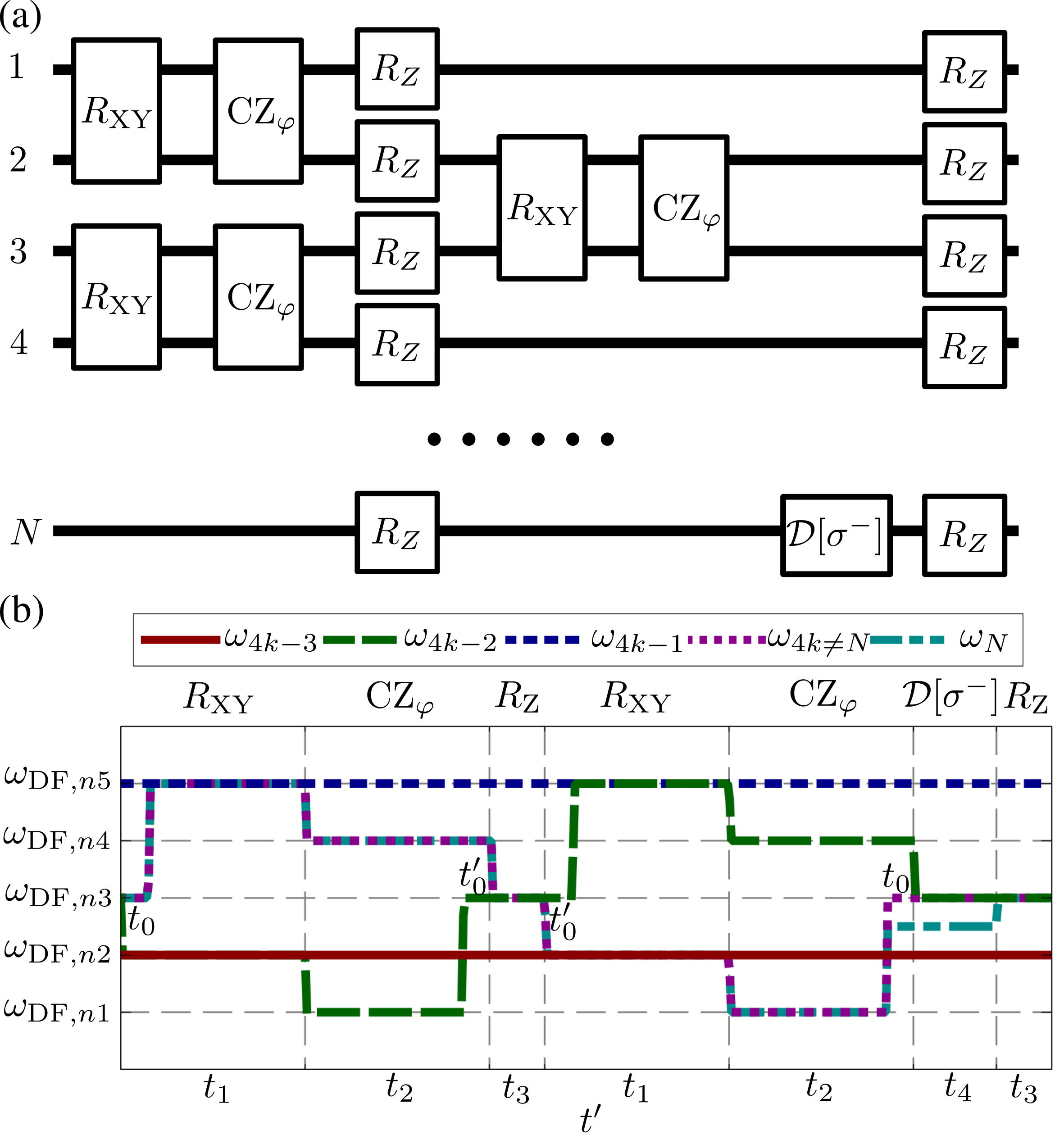}
\caption{Protocol to simulate the dynamics of the dissipative XXZ spin chain using the quantum simulator in \figpanel{fig3}{a}. (a) The operations that need to be performed within a single Trotter step to simulate the dynamics. (b) The protocol to tune the frequencies of the giant atoms to achieve the operations in (a).}
\label{fig5}
\end{figure}

This decomposition enables the simulation of the original model by sequentially applying the dynamics of each component. In particular, $\exp(-iH_1t/l)$ and $\exp(-iH_3t/l)$ correspond to $R_\text{XY}(\theta)$ gates between the corresponding qubits with $\theta=-2Jt/l$. Similarly, $\exp(-iH_2t/l)$ and $\exp(-iH_4t/l)$ yield $R_\text{ZZ}(\varphi_0)$ gates with $\varphi_0=-J_zt/l$. This can be achieved with a CZ$_\varphi$ gate with $\varphi=-4J_zt/l$ and two single-qubit $R_z(-\varphi/4)$ gates. Lastly, $\mathcal{L}_5$ represents single-qubit decay at the chain's end. We note that additional single-qubit $R_z$ gates are required to compensate for phase shifts from frequency tuning before the performance of $R_\text{XY}(\theta)$ gates. Combining all these considerations, the protocol for a single Trotter step is shown in \figpanel{fig5}{a}. 

As detailed in \secref{sec3}, the circuit in \figpanel{fig5}{a} can be implemented by simply adjusting the qubit frequencies. Specifically, the frequencies of odd site qubits are fixed as $\omega_{4k-3}=\omega_{\text{DF},n2}$ and $\omega_{4k-1}=\omega_{\text{DF},n5}$. During the execution of the $R_\text{XY}(\theta)$ gate between qubits $4k-2$ and $4k-3$, the frequency of qubit $4k-2$ is tuned to $\omega_{\text{DF},n2}$, and the system evolves for a duration of $t_1 = \theta/g'_1$, where $g'_1$ denotes the coupling strength between these qubits. Concurrently, the $R_\text{XY}(\theta)$ gate between qubits $4k-1$ and $4k$ is implemented by tuning the frequency of qubit $4k$ to $\omega_{\text{DF},n5}$ and evolving for $t'_1 = \theta/g'_2$, where $g'_2$ represents the coupling strength between qubits $4k-1$ and $4k$. Given that $t'_1 < t_1$, the frequency tuning for qubit $4k$ starts later than that for qubit $4k-2$ by a time interval $t_0 = t_1 - t'_1$ [\figpanel{fig5}{b}]. 

Similarly, other gates in the circuit of \figpanel{fig5}{a} are executed by frequency tuning, enabling the full circuit for a single Trotter step to be realized as shown in \figpanel{fig5}{b}. In this process, $t_2 = \varphi/g'_1$ denotes the duration for executing the CZ gate on qubits $4k-1$ and $4k$, $t'_0 = \varphi/g'_1 - \varphi/g'_2$ represents the time offset between CZ gates on different qubits, $t_3$ is the duration for a single-qubit $R_\text{Z}$ gate, and $t_4 = \Gamma t / (\Gamma_0 l)$ is the simulation time for single-qubit decay, with $\Gamma_0 \approx 1.36\gamma$ being the decay rate to the waveguide when the qubit frequency is set to a value $\omega_\text{decay} \in [\omega_{\text{DF},n2}, \omega_{\text{DF},n3}]$.

\begin{figure}
\center
\includegraphics[width=\linewidth]{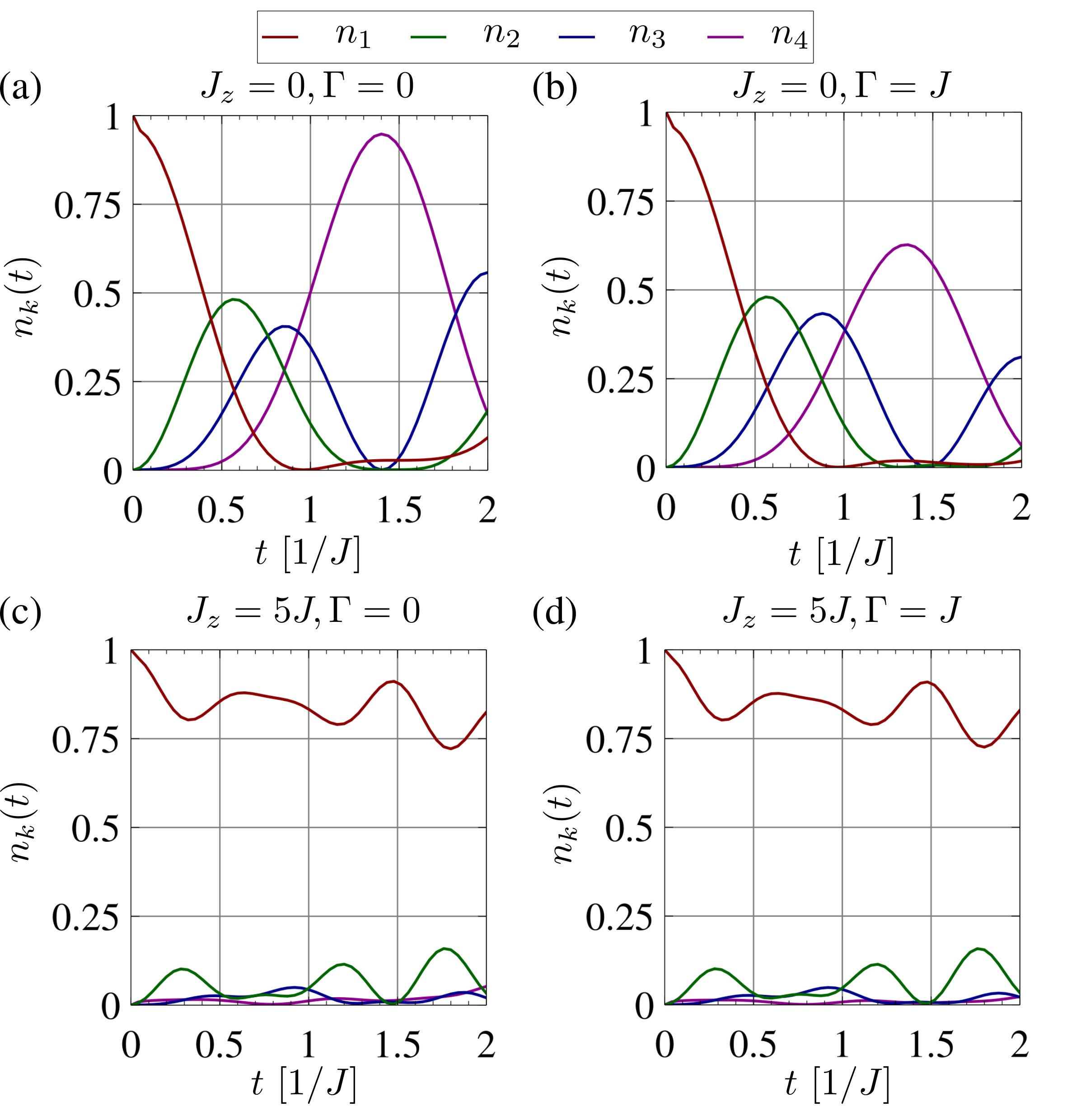}
\caption{Simulation of the dynamics of the dissipative XXZ spin chain [\eqref{eq_model_sim}] with $N=4$ sites, using our giant-atom-based quantum simulator. In (a) and (b), $l=30$ Trotter steps were used; in (c) and (d), $l=10$ Trotter steps were used. The results faithfully capture the slowdown of the spin transport as $J_z$ increases.}
\label{fig6}
\end{figure}

We now present concrete simulation results using typical parameter values: $\gamma/ (2\pi) = \qty{2}{\mega\hertz}$, $\Gamma_\text{ex} = \qty{0.02}{\mega\hertz}$ and $\Gamma_\phi = \qty{0.05}{\mega\hertz}$. $\omega_0$ is set to $\omega_0/(2\pi)=\qty{3.2}{\giga\hertz}$ such that $(\omega_{\text{DF},n2}-\omega_{\text{DF},n1})/(2\pi)\approx\qty{200}{\mega\hertz}$ is achievable as the detuning of the qubits. A conservative single-qubit gate time of $\qty{30}{\nano\second}$ is used~\cite{Blais2021}. The time required for tuning qubit frequencies is negligible ($\sim \qty{1}{\nano\second}$) compared to the simulation duration, as tuning rates of $\sim \qty{0.1}{\giga\hertz/\nano\second}$ are achievable~\cite{Collodo2020}. Thus, we do not include it in our simulations. We simulate the spin dynamics for $N=4$ sites with the initial state $|\psi_0\rangle = \sigma_1^+ |\Omega\rangle$, where $|\Omega\rangle$ is the all-spin-down state. The XXZ Hamiltonian drives spin excitations across the chain, captured by the site populations $n_k(t) = (\langle \psi(t) | \sigma_k^z | \psi(t) \rangle + 1)/2$. This quantity corresponds to the qubit population in the simulator and is experimentally measurable. 

The simulation results are presented in \figref{fig6}, with the number of Trotter steps optimized to balance Trotter and gate errors (see \appref{sec_err}). Without $J_z$ coupling or dissipation, the spin excitation oscillates between sites 1 and 4 [\figpanel{fig6}{a}]. Dissipation at site 4 reduces the oscillation amplitude [\figpanel{fig6}{b}]. The inclusion of $J_z$ interaction decreases spin current, with $J_z = J$ marking the transition between ballistic and diffusive transport regimes~\cite{PhysRevLett.106.217206}. For $J_z = 5J > J$, the reduced spin current diminishes the impact of dissipation on the dynamics [\figpanel{fig6}{d}].

We note that the extended gate set in our simulator minimizes circuit depth compared to alternatives using only iSWAP gates: at least two iSWAP gates (among other single-qubit gates) are required to perform a $R_\text{ZZ}$ gate. Instead, in our case, the $R_\text{ZZ}$ gate is performed with only one CZ$_\varphi$ and single-qubit gates. The efficient compilation provided by our simulator reduces the simulation time and enhances the fidelity of the simulation; as discussed in \secref{sec2}, the increase in fidelity is approximately linear in the decrease in simulation time~\cite{PhysRevLett.129.150504, Abad2025}.



\section{Extension to simulations in higher dimensions and a universal quantum processor} \label{sec5}

The one-dimensional structure of our simulator as it is laid out in \secref{sec3} introduces significant overhead when executing long-range two-qubit gates. Consequently, the implementation of quantum algorithms and protocols requiring such interactions---such as those used in the surface code~\cite{Fowler2012}---becomes inefficient, restricting our simulator’s ability to function as a universal quantum processor. One potential solution to this limitation involves implementing a multi-braided configuration by having the waveguide cross itself once to enable tunable all-to-all couplings between giant atoms~\cite{Chen2025, Kockum2018}. However, this approach introduces a frequency-crowding challenge: distinct two-qubit gates must operate at sufficiently separated frequencies to prevent unwanted interactions during execution. Since qubits function within a constrained frequency range, increasing the number of qubits reduces the available frequency spacing, ultimately limiting scalability.

\begin{figure*}
\center
\includegraphics[width=\linewidth]{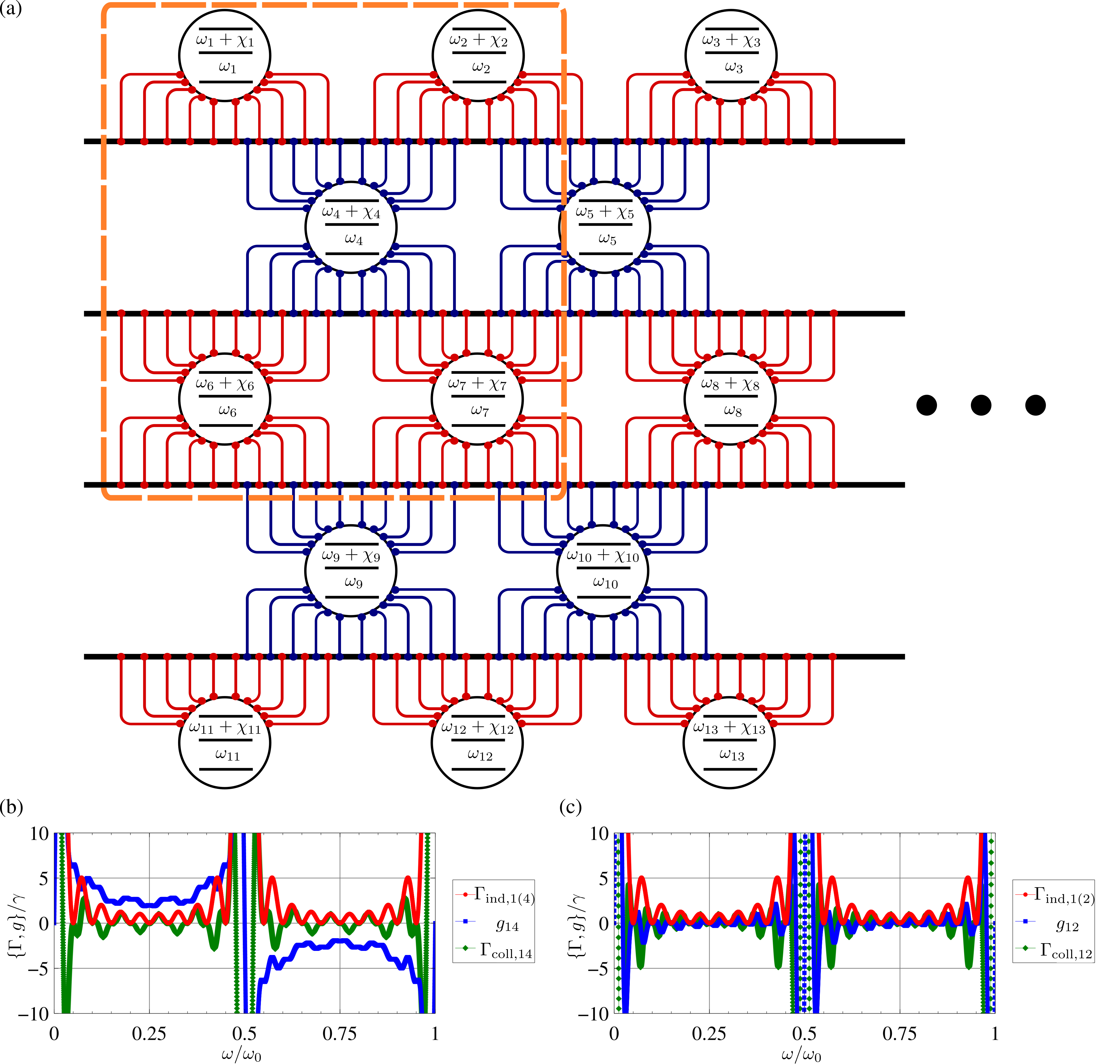}
\caption{A two-dimensional extension of the giant-atom-based simulator from \figpanel{fig3}{a} into a universal quantum processor. (a) Sketch of the structure of the quantum processor. In this design, each giant atom---except those at the boundary---is braided with four neighboring atoms via two waveguides, enabling tunable interactions. This structure supports various two-qubit gate operations through frequency tuning. 
(b,c) The frequency dependence of individual decay rates $\Gamma_\text{ind}$, coupling strengths $g$ and collective decay rates $\Gamma_\text{coll}$ for (b) neighboring and (c) non-neighboring giant atoms.}
\label{fig9}
\end{figure*}

To alleviate this frequency-crowding issue, a square-lattice arrangement~\cite{PhysRevResearch.5.043001, kosen2024signal} can be employed to suppress unwanted interactions between non-neighboring qubits while still enabling both iSWAP and CZ gates. Achieving this with giant atoms necessitates extending our simulator into two dimensions, incorporating an array of waveguides that mediate interactions between different giant atoms, as shown in \figpanel{fig9}{a}. In this setup, the giant atoms (except the ones on the first and last row) are coupled to two waveguides in an identical fashion. This arrangement ensures that the giant atoms can be decoherence-free with respect to both waveguides at decoherence-free frequencies, enabling decoherence-free interactions between giant atoms that allows to execute two-qubit gates. 

Since every giant atom is braided with at most four other giant atoms, we need at least nine decoherence-free frequencies (four for executing $R_{\rm XY}$s, four for executing CZ$_\varphi$, and one for a decoupled regime) within the operational frequency range of the atoms. To minimize individual atom decay within this frequency range, we choose ten coupling points to each waveguide for each giant atom. The coupling points are evenly spaced by $2\Delta x$ and have the same coupling strength $\gamma$. This yields a set of decoherence-free frequencies $\omega_{\text{DF},nm}=(n/2+m/20)\omega_0$ with $n\in\mathcal{N}, m=1,\dots,9$. At $\omega_{\text{DF},nm}$, the braided giant atoms have a non-zero coupling while the unwanted coupling between non-braided giant atoms is eliminated [see \figpanels{fig9}{b}{c}]. Additionally, within the operational frequency regime $[\omega_{\text{DF},n1},\omega_{\text{DF},n9}]$, the individual decay rates of the giant atoms are small, and can further be minimized by optimizing the coupling strengths. Thus, the giant-atom-based quantum processor in \figpanel{fig9}{a} maintains the benefits of the giant-atom-based quantum simulator in \figpanel{fig3}{a}, and furthermore offers possibilities for long-range two-qubit gates owing to its two-dimensional structure.

We now demonstrate how different gates can be executed in the quantum processor in \figpanel{fig9}{a}, focusing on its building block containing five qubits (numbered 1, 2, 4, 6, 7) indicated with the dashed orange rectangle. We set the anharmonicities of the atoms to $\chi_{1,2,6,7}=-\omega_0/20$ to ensure minimal decay. To avoid decoherence and interactions while performing single-qubit gates, the frequencies of the qubits can be set to $\omega_{1(2,4,6,7)}=\omega_{\text{DF},n2(4,5,7,9)}$, such that there are no couplings between the giant atoms. To perform an $R_{\rm XY}$ gate between qubits 1 and 4, we can tune $\omega_4\to\omega_{\text{DF},n2}$; to perform a CZ$_\varphi$ gate between qubits 1 and 4, we can tune $\omega_4\to\omega_{\text{DF},n1}$. Similarly, two-qubit $R_{\rm XY}$ and CZ$_\varphi$ gates can be performed between qubit 4 and other qubits. 

We have thus demonstrated the ability of our two-dimensional quantum processor to execute the extended gate set between neighboring atoms on the square lattice. This in particular allows to perform algorithms such as the surface code~\cite{Andersen2020,Google2023} to enable fault-tolerant quantum computation on our setup, offering it as a universal quantum processor. A detailed analysis of implementing specific algorithms within this structure would require extensive many-body calculations, which we leave for future investigation. 


\section{Discussion}\label{sec6}


We have presented a scalable quantum simulator for open quantum systems. The simulator is based on giant atoms and features an extended gate set that enhances its versatility for simulating open quantum many-body dynamics. Furthermore, we have discussed how to extend the structure in two dimensions for a universal quantum processor. We now move on to discuss (i) the physical realization of the proposed setups and (ii) the potential impact of non-Markovian effects as the system scales up.


\subsection{Physical realization}

A promising platform for implementing our quantum simulator is superconducting qubits~\cite{Gu2017, Blais2021}, such as transmons~\cite{Koch2007}, coupled to a waveguide. There have already been several experiments demonstrating that this platform can be used for giant atoms~\cite{Kannan2020, Vadiraj2021, Joshi2023, Hu2024}. With a typical waveguide speed of light $v\approx 1.3\times \qty[per-mode = symbol]{e8}{\meter\per\second}$ and $\omega_0 / (2\pi) = \qty{3.2}{\giga\hertz}$, the required coupling-point spacing is $\Delta x = 2\pi v/\omega_0\approx\qty{41}{\milli\meter}$. This means that adding a single qubit necessitates an additional waveguide length of $5\Delta x\approx\qty{0.21}{\meter}$. State-of-the-art fabrication techniques can produce waveguides up to $\qty{30}{\meter}$ in length~\cite{Sundaresan2015, Storz2023}, which could accommodate approximately 140 qubits, demonstrating the feasibility of our proposed architecture.

In addition to superconducting qubits, other physical platforms could support the realization of this setup. For example, cold atoms coupled to an optical lattice~\cite{Gonzalez-Tudela2019} present an intriguing alternative, offering distinct advantages in terms of coherence times and system scalability. Exploring such platforms could pave the way for diverse implementations of giant-atom-based quantum simulators.


\subsection{Non-Markovian effects}

Scaling up our simulator enhances non-Markovian effects, which could challenge the validity of \eqref{eq1}. The primary source of non-Markovianity in this system is the time delay associated with photons traveling between coupling points. The Markovian assumption holds as long as $\gamma L_w/v\ll1$, where $L_w$ is the length of the waveguide between coupling points and $\gamma$ is the coupling rate. With $\gamma / (2\pi) = \qty{2}{\mega\hertz}$ and $v = \qty{1.3e8}{\meter/\second}$, this condition is satisfied for $L_w \ll \qty{130}{\meter}$, exceeding current state-of-the-art waveguide lengths.

However, as waveguides approach these lengths or coupling rates increase, deviations from the Markovian regime may emerge. In such cases, incorporating time delays into the theoretical framework and exploring non-Markovian models is necessary. As solving non-Markovian many-body systems remains an open challenge, we leave this for future work. 



\section{Conclusion}\label{sec7}

We have proposed a scalable quantum simulator with an extended gate set, leveraging the unique properties of giant atoms to simulate open quantum systems. The fundamental building block of this processor consists of two giant three-level atoms coupled to the same waveguide in a braided configuration. We demonstrated that this setup enables the realization of both $R_\text{XY}$ and controlled-phase (CZ$_\varphi$) gates through simple frequency tuning of some of the giant atoms. This capability arises from the decoherence-free interaction characteristic of giant atoms, eliminating the need for additional hardware components like parametric couplers.

The scalability of this building block facilitates the construction of a many-body quantum simulator, where nearest-neighbor $R_\text{XY}$ and CZ$_\varphi$ gates can be implemented efficiently by controlling the qubit frequencies. To showcase the simulator’s potential, we performed a Trotterized simulation of the dynamics of a dissipative XXZ spin chain, demonstrating its capability to tackle complex problems in open quantum many-body dynamics.

Our work provides a versatile and scalable platform for quantum simulation, featuring an extended gate set that enhances circuit compilation efficiency while maintaining scalability. The inclusion of both iSWAP-like and controlled-phase gates positions this simulator as a promising candidate for addressing state-of-the-art challenges in quantum simulation, particularly in open quantum many-body physics. Furthermore, extending our simulator into two dimensions could provide a pathway toward a scalable universal quantum processor.

As an outlook for future work, we have already mentioned a detailed analysis of the simulation of specific quantum systems (including ones featuring non-Markovian effects) or of implementation of specific quantum algorithms (including the surface code for error correction), as well as actual experimental implementation with superconducting circuits or other platforms. We also note the possibility of extending the gate set for giant-atom-based simulators even further, e.g., by incorporating three-qubit gates using schemes similar to those in Refs.~\cite{Gu2021, Warren2023}.


\begin{acknowledgements}
    
We thank Liangyu Chen, Akshay Gaikwad, and Laura Garc\'ia \'Alvarez for fruitful discussions. G.C. is supported by European Union's Horizon Europe programme HORIZON-MSCA-2023-PF-01-01 via the project 101146565 (SING-ATOM). A.F.K. acknowledges support from the Swedish Research Council (grant number 2019-03696), the Swedish Foundation for Strategic Research (grant numbers FFL21-0279 and FUS21-0063), the Horizon Europe programme HORIZON-CL4-2022-QUANTUM-01-SGA via the project 101113946 OpenSuperQPlus100, and from the Knut and Alice Wallenberg Foundation through the Wallenberg Centre for Quantum Technology (WACQT).

\end{acknowledgements}


\appendix

\section{Error analysis for the quantum simulation of the dissipative XXZ model} \label{sec_err}



Here we present the numerical details for the simulations presented in \figref{fig6}. The simulator dynamics, which involve 3-level atoms, are governed by the equation
\begin{widetext}
\begin{align} \label{eq_A1}
\partial_t \rho &= -i \mleft[ \sum_{j=1}^N \mleft( \omega_j (\ket{1}\bra{1})_j +  (2\omega_j+\chi_j) (\ket{2}\bra{2})_j \mright) + \sum_{j=1}^N\sum_{k=1}^N g(\omega_j, \omega_{k}) \mleft( \sigma_j^{+,(01)} \sigma_{k}^{-,(01)} + \text{H.c.} \mright) \mright. \nonumber\\
&\quad \mleft. + \sum_{j=1}^N\sum_{k=1}^N \sqrt{2}g(\omega_j + \chi_j, \omega_{k}) \mleft( \sigma_j^{+,(12)} \sigma_{k}^{-,(01)} + \text{H.c.} \mright) + \sum_{j=1}^N\sum_{k=1}^N 2g(\omega_j + \chi_j, \omega_{k} + \chi_k) \mleft( \sigma_j^{+,(12)} \sigma_{k}^{-,(12)}+ \text{H.c.} \mright), \rho \mright] \nonumber \\  
&\quad + \sum_{j=1}^N\Gamma(\omega_j) \mathcal{D}[\sigma_j^{-,(01)}] \rho + \sum_{j=1}^N\sqrt{2}\Gamma(\omega_j + \chi_j) \mathcal{D}[\sigma_j^{-,(12)}] \rho \nonumber \\
&\quad + \sum_{j=1}^N\sum_{k=1}^N\Gamma_\text{coll}(\omega_j,\omega_k) \mleft[ \mleft( \sigma_j^{-,(01)} \rho \sigma_k^{+,(01)} - \frac{1}{2} \mleft\{ \sigma_j^{+,(01)} \sigma_k^{-,(01)} , \rho \mright\} \mright) + \text{H.c.} \mright] \nonumber \\
&\quad + \sum_{j=1}^N\sum_{k=1}^N\sqrt{2}\Gamma_\text{coll}(\omega_j,\omega_k + \chi_k) \mleft[ \mleft( \sigma_j^{-,(01)} \rho \sigma_k^{+,(12)} - \frac{1}{2} \mleft\{ \sigma_j^{+,(01)} \sigma_k^{-,(12)} , \rho \mright\} \mright) + \text{H.c.} \mright] \nonumber \\
&\quad + \sum_{j=1}^N\sum_{k=1}^N\sqrt{2}\Gamma_\text{coll}(\omega_j + \chi_j,\omega_k) \mleft[ \mleft( \sigma_j^{-,(12)} \rho \sigma_k^{+,(01)} - \frac{1}{2} \mleft\{ \sigma_j^{+,(12)} \sigma_k^{-,(01)} , \rho \mright\} \mright) + \text{H.c.} \mright] \nonumber \\
&\quad + \sum_{j=1}^N\sum_{k=1}^N2\Gamma_\text{coll}(\omega_j + \chi_j,\omega_k + \chi_k) \mleft[ \mleft( \sigma_j^{-,(12)} \rho \sigma_k^{+,(12)} - \frac{1}{2} \mleft\{ \sigma_j^{+,(12)} \sigma_k^{-,(12)} , \rho \mright\} \mright) + \text{H.c.} \mright].
\end{align}
\end{widetext}
Here $\omega_j$, $\chi_j$ are the transition frequency and anharmonicity of qubit $j$, respectively, while $\sigma_j^{+,(01)}$ ($\sigma_j^{-,(01)}$) is the raising (lowering) operator of the $\ket {0}$ ($\ket{1}$) level of qubit $j$. The coupling strength between the qubits mediated by the waveguide is denoted by $g$. $\Gamma(\omega_j)$ and $\Gamma_\text{coll}$ represent the individual and collective decay rates of the qubits, respectively, and H.c.~denotes Hermitian conjugate. The parameter dependencies on the qubit frequencies are illustrated in \figpanel{fig3}{b}.

Our simulations operate within the decoherence-free frequency regime, ensuring that non-neighboring qubits do not interact [\figpanel{fig3}{c}]. During the simulation, the time evolution of the qubit frequencies follows the profile shown in \figpanel{fig5}{b}, leading to a time-dependent master equation as in \eqref{eq_A1}, which we solve numerically using QuTiP~\cite{Johansson2012, Johansson2013, Lambert2026}. 

\begin{figure}
\center
\includegraphics[width=\linewidth]{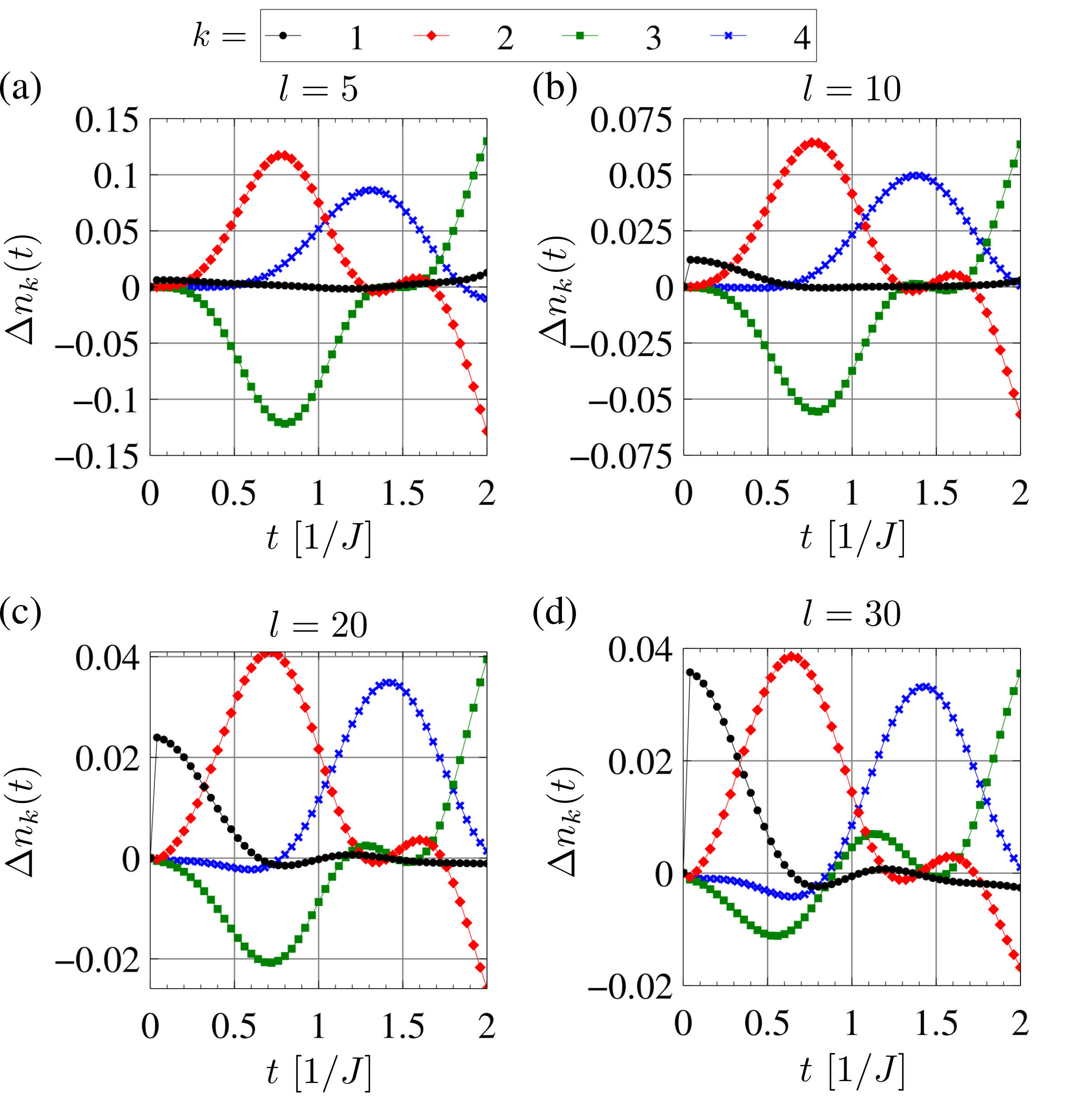}
\caption{Simulation error for the dynamics of the dissipative XXZ spin chain [\eqref{eq_model_sim}] with $J_z=0$, $\Gamma=J$, and $N=4$ sites with our quantum simulator. With a larger number of Trotter steps $l$, the Trotter error is decreased while a larger error stems from the need for more gates to be performed. At short times, gate error dominates and thus a smaller number of Trotter steps results in a smaller error; at long times, Trotter error dominates and a larger number of Trotter steps results in a smaller error.}
\label{figS1}
\end{figure}

We define the simulation error in the simulated population $n_k(t):=(\langle \psi(t)|\sigma_k^z|\psi(t)\rangle+1)/2$ as
\beq
\Delta n_k(t)=\left[ n_k(t) \right]_{\text{exact}}-n_k(t),
\eeq
where $\left[ n_k(t) \right]_{\text{exact}}$ is the exact result. We analyze the simulation error as a function of the number of Trotter steps $l$. 

We first consider the case of $J_z=0$ and $\Gamma=J$ in \eqref{eq_model_sim}, where no CZ$_\varphi$ gates are needed in the simulation. The results are shown in \figref{figS1}. There are two main sources of error: (i) the Trotter error stemming from the Trotterization, which scales as $t^2/l$, and (ii) the gate error, which increases with $l$ since the number of gates grows with $l$. Thus, there is an optimal number of Trotter steps $l_\text{opt}(t)$ for the simulation of the dynamics at time $t$, and in particular, $l_\text{opt}(t)$ increases with $t$. This explanation is in agreement with the simulation error shown in \figref{figS1}, which decreases with $l$ for large $t$, and increases with $l$ for small $t$.

\begin{figure}
\center
\includegraphics[width=\linewidth]{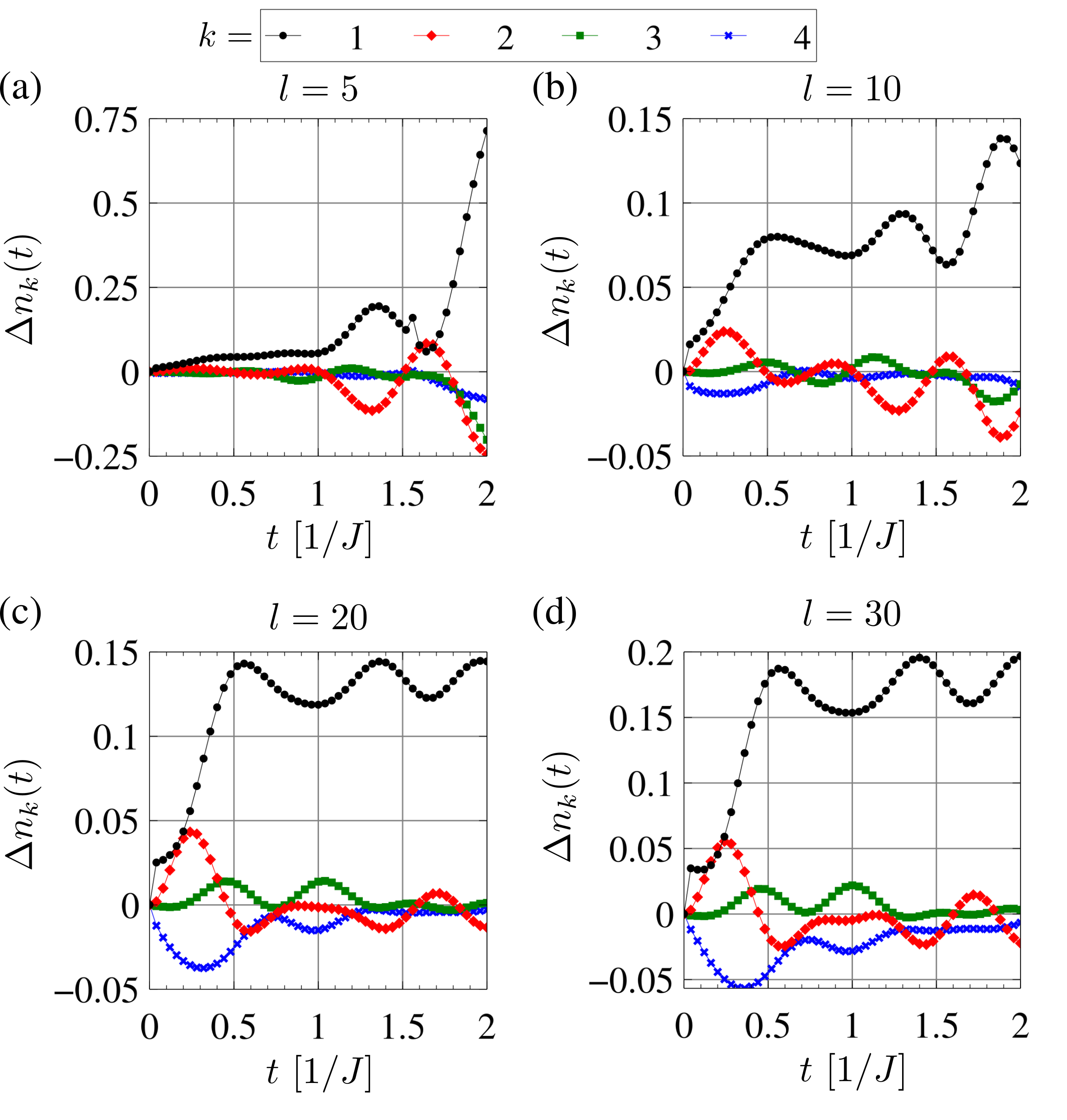}
\caption{Simulation error for the dynamics of the dissipative XXZ spin chain [\eqref{eq_model_sim}] with $J_z=5J$, $\Gamma=J$, and $N=4$ sites with our quantum simulator. Compared to the case of $J_z=0$ in \figref{figS1}, the inclusion of CZ$_\varphi$ gate results in faster growth of gate error with respect to $l$. Thus, the optimal number of Trotter steps has decreased for the same time $t$. In particular, for $t=2$, $l=10$ results in smaller errors than $l=20$.}
\label{figS2}
\end{figure}

We next analyze the simulation error for $J_z=5J$ and $\Gamma=J$ in \eqref{eq_model_sim}. Here, CZ$_\varphi$ gates are necessary, which increases the number of gates per Trotter step. Consequently, the gate error grows more rapidly with $l$ compared to the $J_z=0$ case, resulting in a smaller $l_\text{opt}(t)$.  This is illustrated in \figref{figS2}, where the decreased $l_\text{opt}(t)$ leads to a higher Trotter error and thus greater overall simulation error.

The above analysis underscores the significance of minimizing circuit depth to ensure high simulation accuracy. In particular, a shorter circuit slows the growth of gate error with increasing $l$, enabling a larger $l_\text{opt}(t)$ and reducing simulation error. The extended gate set provided by our simulator effectively reduces circuit depth, enhancing the accuracy of quantum simulations.

\bibliographystyle{quantum}
\bibliography{main,GA,NH_bib}

\end{document}